\documentclass{INTERSPEECH2023}
\usepackage{color,amsmath,url,times,tabularx,bbm,amssymb,multirow}
\usepackage{bm}
\usepackage{cite}
\interspeechcameraready

\title{Improving Audio-Text Retrieval via Hierarchical Cross-Modal Interaction and Auxiliary Captions}
\name{Yifei Xin, Yuexian Zou$^{*}$\thanks{This paper was partially supported by NSFC (No: 62176008) and Shenzhen Science \& Technology Research Program (No:GXWD20201231165807007-20200814115301001).}\thanks{$^{*}$ Yuexian Zou is the corresponding author.}}
\address{School of ECE, Peking University, Shenzhen, China}
\email{xinyifei@stu.pku.edu.cn}

\begin{document}

\maketitle
\begin{abstract}
Most existing audio-text retrieval (ATR) methods focus on constructing contrastive pairs between whole audio clips and complete caption sentences, while ignoring fine-grained cross-modal relationships, e.g., short segments and phrases or frames and words. In this paper, we introduce a hierarchical cross-modal interaction (HCI) method for ATR by simultaneously exploring clip-sentence, segment-phrase, and frame-word relationships, achieving a comprehensive multi-modal semantic comparison. Besides, we also present a novel ATR framework that leverages auxiliary captions (AC) generated by a pretrained captioner to perform feature interaction between audio and generated captions, which yields enhanced audio representations and is complementary to the original ATR matching branch. The audio and generated captions can also form new audio-text pairs as data augmentation for training. Experiments show that our HCI significantly improves the ATR performance. Moreover, our AC framework also shows stable performance gains on multiple datasets.
\end{abstract}

\section{Introduction}
\label{sec:intro}
Given a caption or an audio clip as a query, the audio-text retrieval (ATR) task aims at retrieving a paired item from a set of candidates in another modality. To compute the similarity between the two modalities, a common technique is to embed the whole audio clip and the complete caption sentence into a joint latent space and then adopt a distance metric like the cosine similarity to measure the relevance of the audio and text. However, in the human sense, due to the intrinsic hierarchical semantic structure in audio and text data, we recognize an audio-text pair by simultaneously analyzing audio-text, segment-phrase, and frame-word interactions. Therefore, most existing ATR methods \cite{koepke2022audio,lou2022audio,xin2024diffatr} only considering the single cross-modal interaction between whole audio clips and complete caption sentences would result in a biased retrieval.

In this paper, we introduce a hierarchical cross-modal interaction (HCI) approach for ATR, which hierarchically investigates clip-sentence, segment-phrase, and frame-word interactions to understand audio-text contents comprehensively. To explore fine-grained cross-modal interactions, HCI first constructs hierarchical audio representations and text embeddings at respective frame-segment-clip and word-phrase-sentence granularities, as shown in Figure 1. Taking the audio modality as an illustration, HCI performs attention-based pooling to aggregate semantically correlated frames into several segment representations, which are then fused into a global clip representation. Similar to the audio modality, a sentence also has multi-level representations consisting of words and phrases, which can be expressed in a word-phrase-sentence manner. Thus, based on hierarchical audio and text representations, HCI employs cross-modal contrastive learning to learn inter-modal correlations at frame-word, segment-phrase, and clip-sentence granularities respectively, thereby accomplishing a more comprehensive cross-modal comparison.

Besides, many video-extracted audio clips (e.g., over 2 million audio clips of AudioSet \cite{gemmeke2017audio} are collected from YouTube videos) come with associated text information such as titles and tags, which can be utilized to match textual queries. This motivates us to generate associated captions from audio clips to benefit ATR. To achieve this, we present a novel ATR framework that utilizes the pretrained audio captioner (e.g., the widely used CNN10 audio encoder from PANNs \cite{kong2020panns} and the GRU decoder \cite{xu2021investigating}) to generate captions for each audio clip. The generated captions can be utilized from three aspects. First, the provided audio clip and its generated caption are a matched pair, so they can be used as extra positive sample pairs in addition to the initial audio-text pair as data augmentation during the training stage. Second, we perform cross-modal interaction between the audio and generated captions to enhance audio features. Specifically, we can make use of the information complementarity between audio clips and captions to reduce redundant features from audio clips and learn more discriminative audio representations. Third, we can leverage the text-caption matching to complement the original text-audio matching for ATR, thus reducing the bias of the model and yielding more robust retrieval results.

\begin{figure}[t]
  \centering
  \includegraphics[width=1.0\linewidth]{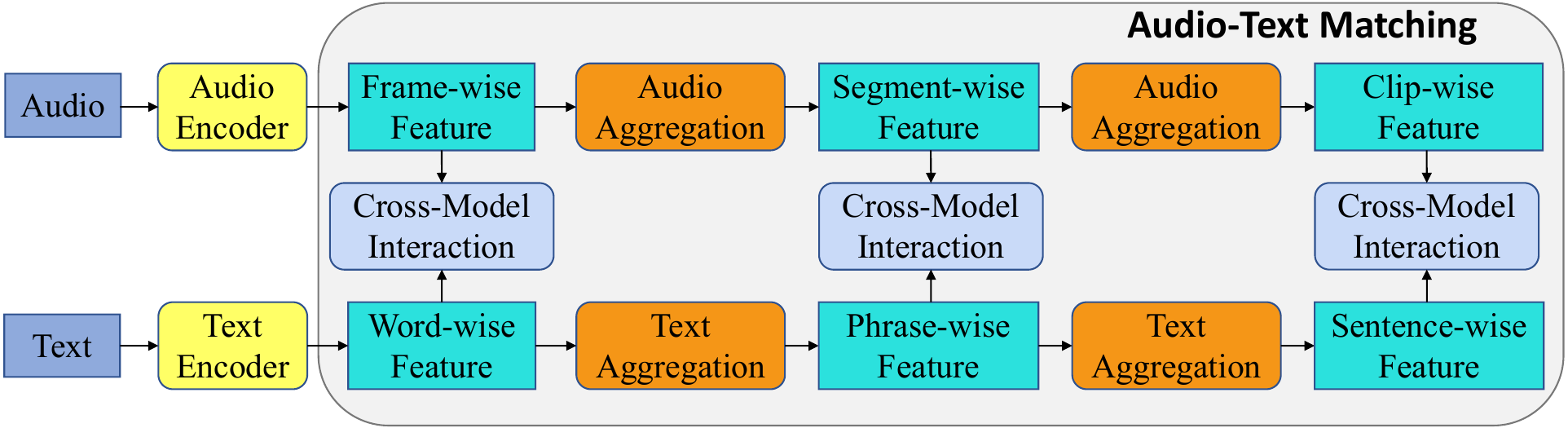}
  \caption{The overview of our hierarchical cross-modal interaction method for ATR.}
  \label{fig:adapt-pic}
  \vspace*{-\baselineskip}
  \vspace*{-0.2cm}
\end{figure}

In a nutshell, our contributions are threefold: 
\begin{itemize}
\item We introduce a hierarchical cross-modal interaction method for ATR, which explores multi-level cross-modal interactions at clip-sentence, segment-phrase, and frame-word granularities to understand audio-text contents comprehensively. 
\item We present a novel ATR framework that makes full use of the generated captions from three aspects (i.e., data augmentation, feature interaction, complementary text-caption matching) to produce more robust retrieval results.
\item Experiments show that our HCI effectively improves the ATR performance. Moreover, our AC framework also shows stable performance gains on multiple datasets.
\end{itemize}
\vspace*{-0.1cm}
\section{Problem Formulation}
\label{sec:method}
Let $D = \lbrace(a_i, t_i)\rbrace_{i=1}^N$ be an audio retrieval dataset containing $N$ samples, where $a_i$ is an audio clip and $t_i$ is the paired text. Therefore, $(a_i, t_i)$ is regarded as a positive pair while $(a_i, t_j,j \neq i)$ is a negative pair. The ATR models usually consist of a text encoder $f_t(\cdot)$ and an audio encoder $f_a(\cdot)$ pretrained on the sound event detection task \cite{mesaros2021sound,xin2023background,xin2023enhancement}, which project the text and audio into a shared embedding space, respectively. For an audio-text pair $(a_i, t_j)$, the similarity of the audio and text can be measured by the cosine similarity of their embeddings:
\begin{equation}
             s(a_i, t_i) = \frac{f_a(a_i) \cdot f_t(t_i)}{{\Vert f_a(a_i) \Vert}_2 {\Vert f_t(t_i) \Vert}_2}.
\end{equation}
Currently, the NT-Xent loss \cite{mei2022metric,chen2020simple} based on symmetrical cross-entropy is widely employed, which has been shown to consistently outperform the previous triplet-based losses \cite{xie2022dcase,lamorttake}. Therefore, we adopt it as the baseline loss function for our work. The NT-Xent loss is formulated as below:
\begin{equation}
\begin{aligned}
    \mathcal{L}_{at} = -\frac{1}{N} \left(\sum_i^N {\rm log}\frac{{\rm exp}(s(a_i, t_i)/\tau)}{\sum_j^N {\rm exp}(s(a_i, t_j)/\tau)}+ \right.\ \\ \left. \sum_i^N {\rm log}\frac{{\rm exp}(s(a_i, t_i)/\tau)}{\sum_j^N {\rm exp}(s(a_j, t_i)/\tau)} \right),
\end{aligned}
\end{equation}
where $\tau$ is a temperature hyper-parameter for scaling. Following the previous work \cite{mei2022metric}, we set $\tau$ = 0.07 in our experiments. The training objective is to maximize the similarity of the positive pair relative to all negative pairs, and the ultimate loss is calculated in both directions.
 \begin{figure}[t]
  \centering
  \includegraphics[width=1.0\linewidth]{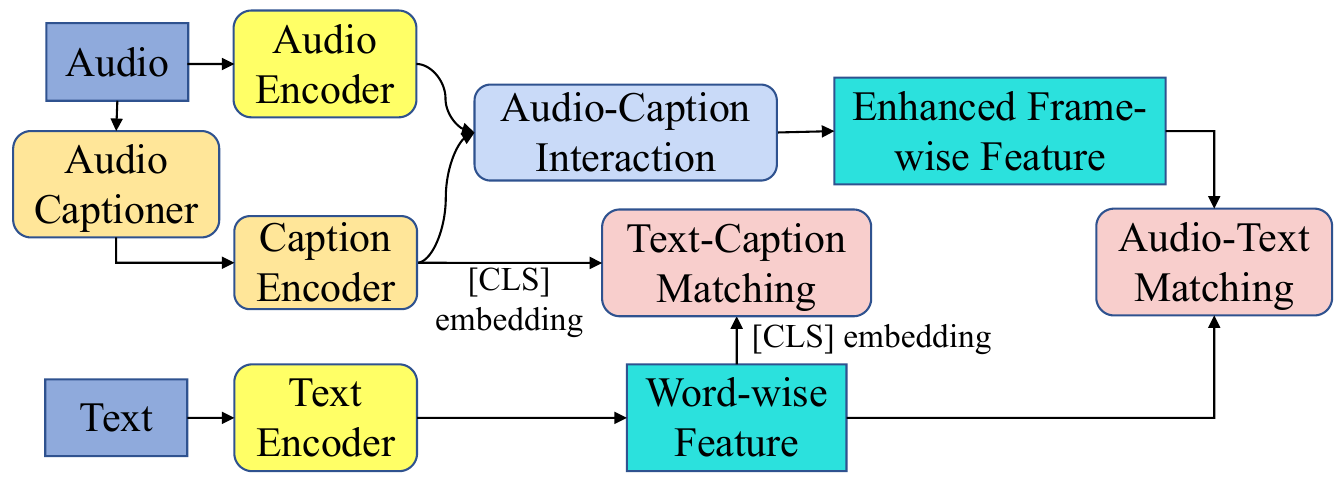}
  \caption{The overview of our auxiliary captions (AC) framework for ATR.}
  \label{fig:adapt-pic}
  \vspace*{-\baselineskip}
  \vspace*{-0.2cm}
\end{figure}
\vspace*{-0.1cm}
\section{Proposed methods}
\label{sec:proposed method}
\subsection{Hierarchical Cross-Modal Interaction}
We denote $A_i^f \in \mathbb{R}^{N_f \times D}$ as the frame representations extracted from the audio encoder, and $T_i^w \in \mathbb{R}^{N_w \times D}$ as the word embeddings extracted from the text encoder, where $N_f$ is the number of audio frames, $N_w$ is the number of words, and $D$ is the feature dimension. To further extract features that capture temporal audio information and long-term word dependence, HCI leverages self-attention \cite{touvron2021augmenting,wang2019comparison,jiang2022tencent} to aggregate semantically related frames into segment representations and related words into phrase embeddings. Taking audio modality as an example, the aggregation function $g_a(\cdot)$ is defined as:
\begin{equation}
             A_i^s = g_a(A_i^f) = {\rm softmax}(A_i^f W_s)^T h(A_i^f),
\end{equation}
where $W_s \in \mathbb{R}^{D \times N_s}$ ($N_s$ is the number of audio segments), and ${\rm softmax}(A_i^f W_s)^T$ projects $A_i^f$ into normalized frame weights with the dimension of ${N_s \times N_f}$. $h(\cdot)$ is a two-layer FC-ReLU with feature channel changes $D$-$2D$-$D$. Thus, $g_a(\cdot)$ aggregates frame representations $A_i^f$ into segment representations $A_i^s \in \mathbb{R}^{N_s \times D}$, where $N_s < N_f$. We denote $A_i^s = \lbrace A_{i,1}^s,...,A_{i,N_s}^s \rbrace$ as a set of $N_s$ audio segments. $A_{i,j}^s$ aggregates several semantically related frame representations into a single one, which contains segment information. Similarly, the text aggregation function $g_t(\cdot)$ is given as:
\begin{equation}
             T_i^p = g_t(T_i^w) = {\rm softmax}(T_i^w W_p)^T h(T_i^w),
\end{equation}
where $W_p \in \mathbb{R}^{D \times N_p}$ ($N_p$ is the number of phrases), and $T_i^p \in \mathbb{R}^{N_p \times D}$. $A_i^s$ and $T_i^p$ can be further aggregated into the clip-level representation $A_i^c \in \mathbb{R}^{1 \times D}$ and the sentence-level embedding $T_i^s \in \mathbb{R}^{1 \times D}$ using $g_a(\cdot)$ and $g_t(\cdot)$. Based on $\lbrace A_i^f, A_i^s, A_i^c \rbrace$ and $\lbrace T_i^w, T_i^p, T_i^s \rbrace$, HCI can perform comprehensive semantic comparison of audio-text pairs at frame-segment-clip and word-phrase-sentence granularities, respectively. 

For $\lbrace A_i^f, T_i^w \rbrace$, the contrastive loss of frame-word interaction $\mathcal{L}_{f-w}$ and the cross-model interaction (CI) function are given by:
\begin{equation}
\begin{aligned}
    \mathcal{L}_{f-w} = -\frac{1}{N} \left(\sum_i^N {\rm log}\frac{{\rm exp}(CI(A_i^f, T_i^w)/\tau)}{\sum_j^N {\rm exp}(CI(A_i^f, T_j^w)/\tau)}+ \right.\ \\ \left. \sum_i^N {\rm log}\frac{{\rm exp}(CI(A_i^f, T_i^w)/\tau)}{\sum_j^N {\rm exp}(CI(A_j^f, T_i^w)/\tau)} \right),
\end{aligned}
\end{equation}
\vspace*{-0.1cm}
\begin{equation}
\begin{aligned}
   CI(A_i^f, T_i^w) = \left(\frac{1}{N_w} \sum_{n=1}^{N_w} {\rm max_{m=1}^{N_f}} s(A_{i,m}^f, T_{i,n}^w) + \right.\ \\ \left. \frac{1}{N_f} \sum_{n=1}^{N_f} {\rm max_{m=1}^{N_w}} s(A_{i,n}^f, T_{i,m}^w) \right) / 2,
\end{aligned}
\end{equation}
where $s(\cdot, \cdot)$ denotes the cosine similarity of the two embeddings. $CI(A_i^f, T_i^w)$ first computes a pair-wise similarity matrix between frames and words and then aggregates all frame-word similarities into an overall score. $L_{f-w}$ is a symmetric cross-modal contrastive loss that measures the cross-modal similarity between a set of frames and words.

Similarly, for $\lbrace A_i^s, T_i^p \rbrace$, the contrastive loss of segment-phrase interaction $\mathcal{L}_{s-p}$ and the CI module are denoted as:
\begin{equation}
\begin{aligned}
    \mathcal{L}_{s-p} = -\frac{1}{N} \left(\sum_i^N {\rm log}\frac{{\rm exp}(CI(A_i^s, T_i^p)/\tau)}{\sum_j^N {\rm exp}(CI(A_i^s, T_j^p)/\tau)}+ \right.\ \\ \left. \sum_i^N {\rm log}\frac{{\rm exp}(CI(A_i^s, T_i^p)/\tau)}{\sum_j^N {\rm exp}(CI(A_j^s, T_i^p)/\tau)} \right),
\end{aligned}
\end{equation}
\vspace*{-0.1cm}
\begin{equation}
\begin{aligned}
   CI(A_i^s, T_i^p) = \left(\frac{1}{N_s} \sum_{n=1}^{N_s} {\rm max_{m=1}^{N_p}} s(A_{i,m}^s, T_{i,n}^p) + \right.\ \\ \left. \frac{1}{N_p} \sum_{n=1}^{N_p} {\rm max_{m=1}^{N_s}} s(A_{i,n}^s, T_{i,m}^p) \right) / 2.
\end{aligned}
\end{equation}

For the cross-model interaction between $\lbrace A_i^c, T_i^s \rbrace$, $\mathcal{L}_{c-s}$ is actually the original NT-Xent loss, which uses cosine similarity to measure the cross-modal similarity between the global clip and sentence representations as the baseline method does. Note that following the previous work \cite{mei2022metric}, we use the [CLS] text embedding \cite{mei2022language} for the global sentence representation, as it performs better than the sentence-level embedding aggregated from the word-level embeddings.

Finally, the loss function for our HCI is:
\begin{equation}
\begin{aligned}
    \mathcal{L}_{hci} = \mathcal{L}_{c-s} + \alpha \mathcal{L}_{f-w} + \beta \mathcal{L}_{s-p},
\end{aligned}
\end{equation}
where $\alpha$ and $\beta$ are the hyper-parameters.

\subsection{Auxiliary Captions}
In this section, we first introduce the way to generate associated captions for audio clips. Then, we detail how to make full use of the generated captions to improve the ATR performance. Figure 2 illustrates the overview of our auxiliary caption framework.

\subsubsection{Caption Generation}
To get the auxiliary caption for the given audio clip, we utilize the pretrained  encoder-decoder model to generate correlated captions. Specifically, we utilize a 10-layer CNN (CNN10) \cite{kong2020panns} as the audio encoder and a standard shallow single layer unidirectional GRU \cite{xu2021investigating} as the decoder, which are all commonly used on the audio captioning task \cite{zhang2022caption,mei2022automated,zhou2022can}. Both the encoder and decoder pretrained on the audio captioning dataset are frozen when generating captions for our auxiliary captions (AC) framework. Besides, we adopt the BERT \cite{kenton2019bert} as our caption encoder to generate the caption embedding.

\subsubsection{Data Augmentation with Auxiliary Captions}
Given the generated captions, the most obvious use for the auxiliary captions is to augment training data. For instance, given a dataset including $N$ audio clips and corresponding texts, each audio clip and its generated caption are a matched pair, so they can be treated as extra positive sample pairs in addition to the audio-text pair for training. As a result, we can increase $N$ pairs as additional data augmentation during the training stage.
\begin{table}\footnotesize
  \caption{Performance comparison of our HCI method with $L_{hci}$ and previous methods with the NT-Xent loss.}
  \centering
  \vspace*{-\baselineskip}
  \label{tab:freq}
  \begin{tabular}{c|cc|cc}
    \toprule
    \multirow{2}{*}{Methods} & \multicolumn{2}{c|}{Text-to-Audio} & \multicolumn{2}{c}{Audio-to-Text} \\
    & \textbf{R@1} & \textbf{R@10} & \textbf{R@1} & \textbf{R@10}\\
    \midrule
    \multicolumn{5}{c}{\textbf{AudioCaps}}\\
    \midrule
    ResNet38+NTXent \cite{mei2022metric} & 33.9 & 82.6 & 39.4 & 83.9\\
    CNN14+NTXent & 31.4 & 78.8 & 38.2 & 81.9\\   
    \textbf{ResNet38+HCI} & \textbf{36.6} & \textbf{85.6} & \textbf{41.9} & \textbf{85.8}\\
    \textbf{CNN14+HCI} & \textbf{33.9} & \textbf{81.4} & \textbf{41.1} & \textbf{84.3}\\
    \midrule
    \multicolumn{5}{c}{\textbf{Clotho}}\\
    \midrule
    ResNet38+NTXent \cite{mei2022metric} & 14.4 & 49.9 & 16.2 & 50.2\\
    CNN14+NTXent & 13.9 & 48.2 & 14.3 & 49.9\\   
    \textbf{ResNet38+HCI} & \textbf{16.8} & \textbf{52.8} & \textbf{19.1} & \textbf{52.7}\\
    \textbf{CNN14+HCI} & \textbf{15.9} & \textbf{50.1} & \textbf{16.2} & \textbf{51.6}\\
  \bottomrule
\end{tabular}
\vspace*{-\baselineskip}
\vspace*{-0.2cm}
\end{table}

\subsubsection{Audio-Caption Cross-Modal Interaction}
We also perform cross-modal interactions between the audio and generated caption to enhance audio representations. Our motivation is to make use of the information complementarity between audio clips and captions to reduce redundant features from audio and learn more discriminative audio representations. Specifically, we feed the frame-level audio embeddings (as query) and the [CLS] embedding of the generated caption (as key and value) into the audio-caption interaction module. The audio-caption interaction module employs one co-attention transformer layer \cite{lu2019vilbert,chen2021multimodal,wu2022cap4video} to facilitate cross-modality information exchange, which passes the keys and values (caption embeddings) from the caption modality to the queries (frame-level embeddings) of the audio modality, followed by a standard transformer layer to model temporal information, thereby obtaining enhanced frame-wise audio representations.
\vspace*{-0.1cm}
\subsubsection{Complementary Text-Caption Matching}
In addition to the uses of captions for data augmentation and audio representation enhancement mentioned above, the generated caption itself can also reflect the content of the audio, allowing us to leverage the generated caption for text-caption retrieval. Specifically, each caption generated by the audio clip is then passed through the caption encoder to obtain its [CLS] embedding. Then, the cosine similarity between the caption embedding and the text embedding is calculated to complement the audio-text matching.

We denote $T_i^c$, $C_i^c$ as the text and the caption [CLS] embeddings of the $i$-th sample, respectively. For the text-caption branch, it is preferable that the text embedding $T_i^c$ and the caption embedding $C_i^c$ be close when they are related and far apart when they are not during the training stage. We follow the common practice \cite{mei2022metric,chen2020simple,xin2024audio} to consider the bidirectional learning objective, which uses the symmetric cross-entropy loss to maximize the similarity between matched text-caption pairs while minimizing the similarity for other pairs:
\begin{equation}
\begin{aligned}
    \mathcal{L}_{tc} = -\frac{1}{N} \left(\sum_i^N {\rm log}\frac{{\rm exp}(s_{tc}(T_i^c, C_i^c)/\tau)}{\sum_j^N {\rm exp}(s_{tc}(T_i^c, C_j^c)/\tau)}+ \right.\ \\ \left. \sum_i^N {\rm log}\frac{{\rm exp}(s_{tc}(T_i^c, C_i^c)/\tau)}{\sum_j^N {\rm exp}(s_{tc}(T_j^c, C_i^c)/\tau)} \right),
\end{aligned}
\end{equation}
where $s_{tc}(\cdot, \cdot)$ represents the text-caption matching similarity function. The total loss $\mathcal{L}_{total}$ is the sum of audio-text loss $\mathcal{L}_{at}$ and the text-caption loss $\mathcal{L}_{tc}$:
\begin{equation}
\begin{aligned}
    \mathcal{L}_{total} = \mathcal{L}_{at} + \mathcal{L}_{tc}.
\end{aligned}
\end{equation}
Notably, the $\mathcal{L}_{at}$ can be replaced by our HCI loss $\mathcal{L}_{hci}$, which can achieve a higher performance.
\begin{table}\footnotesize
  \caption{Performance comparison of our auxiliary captions (AC) framework with previous methods.}
  \centering
  \vspace*{-\baselineskip}
  \label{tab:freq}
  \begin{tabular}{c|cc|cc}
    \toprule
    \multirow{2}{*}{Methods} & \multicolumn{2}{c|}{Text-to-Audio} & \multicolumn{2}{c}{Audio-to-Text} \\
    & \textbf{R@1} & \textbf{R@5} & \textbf{R@1} & \textbf{R@5}\\
    \midrule
    \multicolumn{5}{c}{\textbf{AudioCaps}}\\
    \midrule
    Audio-Text \cite{mei2022metric} & 33.9 & 69.7 & 39.4 & 72.0\\
    \textbf{Audio-Text+AC} & \textbf{35.4} & \textbf{71.3} & \textbf{41.1} & \textbf{73.6}\\
    \textbf{Audio-Text+AC+HCI} & \textbf{37.2} & \textbf{72.7} & \textbf{43.3} & \textbf{75.2}\\
    \midrule
    \multicolumn{5}{c}{\textbf{Clotho}}\\
    \midrule
    Audio-Text \cite{mei2022metric} & 14.4 & 36.6 & 16.2 & 37.5\\
    \textbf{Audio-Text+AC} & \textbf{16.8} & \textbf{38.7} & \textbf{18.7} & \textbf{39.9}\\
    \textbf{Audio-Text+AC+HCI} & \textbf{18.2} & \textbf{39.4} & \textbf{19.9} & \textbf{41.3}\\
  \bottomrule
\end{tabular}
\vspace*{-\baselineskip}
\vspace*{-0.2cm}
\end{table}
\section{Experiments and Results}
\label{sec:exp}
\subsection{Datasets}
We evaluate our methods on two publicly available datasets: AudioCaps \cite{kim2019audiocaps} and Clotho \cite{drossos2020clotho} datasets. AudioCaps contains about 50K audio samples, which are all 10-second long. The training set consists of 49274 audio clips, each with one human-annotated caption. The validation and test sets contain 494 and 957 audio clips, each with five human-annotated captions. The Clotho v2 dataset contains 6974 audio samples between 15 and 30 seconds in length. Each audio sample is annotated with 5 sentences. The numbers of training, validation, and test samples are 3839, 1045, and 1045, respectively. 

\subsection{Training Details and Evaluation metrics}
In our work, we follow the same pipeline in \cite{mei2022metric} to train our networks. We adopt BERT \cite{kenton2019bert} as the text encoder, while employing the ResNet-38 and CNN14 in pre-trained audio neural networks (PANNs) \cite{kong2020panns} as the audio encoder. The dataset for captioner pretraining is consistent with the dataset for the ATR task (e.g., the captioner are pretrained on the AudioCaps when training and evaluating the retrieval performance on the AudioCaps dataset). We conduct experiments by fine-tuning the pretrained text and audio encoders, while freezing the pretrained captioner and the caption encoder to generate the caption and its embedding for our AC framework. The hyper-parameters are set as $N_s$ = $N_p$ = 10, $\alpha$ = 0.5, $\beta$ = 0.1. Recall at rank k (R@k) is utilized as the evaluation metric, which is a popular cross-modal retrieval evaluation protocol \cite{xin2023cooperative,zhao2024mint,cheng2023s}. R@k measures the proportion of targets retrieved within the top-k ranked results, so a higher score means better performance. The results of R@1, R@5, and R@10 are reported.

\subsection{Experimental Results}
As shown in Table 1, we first compare the performance of our hierarchical cross-modal interaction (HCI) method (using $\mathcal{L}_{hci}$) with previous baselines (using the NT-Xent loss). We adopt either the ResNet-38 or the CNN14 as the audio encoder on the AudioCaps and Clotho datasets. It can be seen that our HCI brings significant gains with different audio encoders on both datasets, thus demonstrating the effectiveness of our method. 

To evaluate our auxiliary captions (AC) framework, we compare it with the baseline method of just using original audio-text pairs to compute the similarity with the NT-Xent loss. Here, we choose ResNet-38 as the audio encoder. As can be seen in Table 2, our AC framework also achieves performance boosts by a large margin. Moreover, when using our $L_{hci}$ to replace the NT-Xent loss, the performance can be further improved, which strongly proves the effectiveness and robustness of our HCI method and AC framework.

\begin{table}\small
  \caption{Ablation study of our $\mathcal{L}_{hci}$ loss.}
  \centering
  \vspace*{-\baselineskip}
  \label{tab:freq}
  \begin{tabular}{c|cc|cc}
    \toprule
    \multirow{2}{*}{Methods} & \multicolumn{2}{c|}{Text-to-Audio} & \multicolumn{2}{c}{Audio-to-Text} \\
    & \textbf{R@1} & \textbf{R@10} & \textbf{R@1} & \textbf{R@10}\\
    \midrule
    $\mathcal{L}_{c-s}$ \cite{mei2022metric} & 33.9 & 82.6 & 39.4 & 83.9\\
    \midrule
    +$\mathcal{L}_{f-w}$ & 35.9 & 85.1 & 41.1 & 85.2\\       
    +$\mathcal{L}_{s-p}$ & 34.4 & 84.3 & 40.6 & 84.6\\   
    +$\mathcal{L}_{f-w}$+$\mathcal{L}_{s-p}$ & \textbf{36.6} & \textbf{85.6} & \textbf{41.9} & \textbf{85.8}\\   
  \bottomrule
\end{tabular}
\vspace*{-\baselineskip}
\end{table}

\begin{table}
  \caption{Influences of different text embeddings, and segment-phrase numbers on the AudioCaps dataset.}
  \vspace*{-\baselineskip}
  \centering
  \label{tab:freq}
  \begin{tabular}{c|cc|cc}
    \toprule
    \multirow{2}{*}{Methods} & \multicolumn{2}{c|}{Text-to-Audio} & \multicolumn{2}{c}{Audio-to-Text} \\
    & \textbf{R@1} & \textbf{R@10} & \textbf{R@1} & \textbf{R@10}\\
    \midrule
    HCI (Avg) & 36.1 & 84.9 & 41.4 & 85.2\\
    \textbf{HCI ([CLS])} & \textbf{36.6} & \textbf{85.6} & \textbf{41.9} & \textbf{85.8}\\
    \midrule
    HCI ($N_s$=8) & 36.3 & 85.5 & 41.5 & 85.3\\
    \textbf{HCI ($N_s$=10)} & \textbf{36.6} & \textbf{85.6} & \textbf{41.9} & \textbf{85.8}\\
    HCI ($N_s$=12) & 36.2 & 85.2 & 41.4 & 85.6\\
  \bottomrule
\end{tabular}
\vspace*{-\baselineskip}
\vspace*{-0.2cm}
\end{table}

\subsection{Ablation Study}
In this part, we discuss the influence of each term of our $\mathcal{L}_{hci}$ loss, the selection of text embeddings with the [CLS] token and the aggregated sentence-level embedding, different segment-phrase numbers, and each component of our AC framework. Here, we use the ResNet-38 as the audio encoder.
\begin{table}\small
  \caption{Influences of each part of our AC framework.}
  \centering
  \vspace*{-\baselineskip}
  \label{tab:freq}
  \begin{tabular}{c|cc|cc}
    \toprule
    \multirow{2}{*}{Methods} & \multicolumn{2}{c|}{Text-to-Audio} & \multicolumn{2}{c}{Audio-to-Text} \\
    & \textbf{R@1} & \textbf{R@5} & \textbf{R@1} & \textbf{R@5}\\
    \midrule
    Audio-Text \cite{mei2022metric} & 14.4 & 36.6 & 16.2 & 37.5\\
    \midrule
    +DA & 15.1 & 37.2 & 16.8 & 38.2\\       
    +ACFI & 15.9 & 37.8 & 17.6 & 39.1\\   
    \textbf{+TCM} & \textbf{16.8} & \textbf{38.7} & \textbf{18.7} & \textbf{39.9}\\   
  \bottomrule
\end{tabular}
\vspace*{-\baselineskip}
\vspace*{-0.25cm}
\end{table}

\textbf{Results of each term of our $\mathcal{L}_{hci}$ loss.} To show the effectiveness of each term of our $\mathcal{L}_{hci}$ loss, we present ablation results on the AudioCaps dataset in Table 3, where $\mathcal{L}_{c-s}$ is our baseline method using the NT-Xent loss. It can be seen that compared to the segment-phrase interaction, the frame-word interaction brings a larger performance benefit, while combining the frame-word and segment-phrase interactions can further boost the performance, indicating that performing the multi-level interaction between audio and text can effectively benefit the ATR task.

\textbf{Results of different text embeddings, and segment-phrase numbers.} As shown in Table 4, HCI (*) means that we only change the setting of *, and the rest of the settings are the same as the experimental best settings, which we show in bold. We first conduct experiments for comparing the performance of the [CLS] text embedding and the aggregated sentence-level embedding. It is clear that the [CLS] embedding consistently performs better than the aggregated sentence-level embedding. Then, we experiment with different $N_s$ (= $N_p$) numbers, indicating how many segments or phrases HCI obtains from frames or words. It is clear that $N_s$ = $N_p$ = 10 achieves the best performance. 

\textbf{Results of each part of our AC framework.} As shown in Table 5, we provide ablation results to clarify the effects of our AC framework on the Clotho dataset, where +DA represents that we use auxiliary captions only for data augmentation based on our baseline models, +ACFI denotes that we add the audio-caption interaction module based on +DA, and +TCM means to add the text-caption matching module based on +ACFI. It can be seen that our method achieves consistent performance gains by merging each component of our AC framework step by step, which highly verifies the effectiveness of our AC framework. 

\section{Conclusions}
\label{sec:conclusion}
In this paper, we present a hierarchical cross-modal interaction (HCI) method for ATR, which simultaneously explores clip-sentence, segment-phrase, and frame-word relationships to understand audio-text contents. Besides, we also develop a framework that leverages auxiliary captions (AC) generated by a pretrained captioner to benefit the audio-text matching from three aspects, i.e., data augmentation, audio-caption feature interaction for enhancing audio representations, and text-caption matching to complement the original ATR matching branch. Experiments show that our HCI significantly improves the ATR performance. Moreover, our AC framework also shows stable performance gains on the AudioCaps and Clotho datasets.

\bibliographystyle{IEEEtran}
\bibliography{mybib}

\end{document}